\documentclass[english,twocolumn,aps,superscriptaddress,pra,10pt,floatfix]{revtex4-2}
\usepackage{color}
\usepackage{xcolor}
\usepackage{amsmath}
\usepackage{amssymb}
\usepackage{graphicx}
\usepackage{array}
\usepackage{multirow}
\usepackage{makecell}
\definecolor{ForestG}{rgb}{0.31, 0.78, 0.47}
\definecolor{MyGreen}{RGB}{54,165,54}

\makeatletter
\usepackage{amsmath}
\usepackage{amssymb}
\usepackage{inputenc}
\nopagebreak

\usepackage{babel}
\begin{document}
\title{Composition dependent $\mathbf{k}\cdot\mathbf{p}$ band parameters for wurtzite  (Al,Ga)N alloys from
density functional theory}
\author{Amit Kumar Singh}
\affiliation{Tyndall National Institute, University College Cork, Dyke Parade, Cork T12 R5CP, Ireland}
\author{Alvaro Gomez-Iglesias}
\affiliation{ams-OSRAM International GmbH, Regensburg 93055, Germany}
\author{Stefan Schulz }
\email{stefan.schulz@tyndall.ie}

\affiliation{Tyndall National Institute, University College Cork, Dyke Parade, Cork T12 R5CP, Ireland}
\affiliation{School of Physics, University College Cork, Cork T12 YN60, Ireland}

\begin{abstract}
Ultraviolet light emitters based on the semiconductor alloy aluminium gallium nitride, (Al,Ga)N, have attracted significant interest in recent years due to their potential for optoelectronic devices. To guide the design of such devices with improved efficiencies, theoretical frameworks based on so-called $\mathbf{k}\cdot\mathbf{p}$ methods have found widespread application in the literature. Given that $\mathbf{k}\cdot\mathbf{p}$ models are empirical in nature, parameters such as effective masses or crystal field splitting energies of (Al,Ga)N alloys have to be provided as input from first-principles calculations or experiment. Although these parameters are available for GaN and AlN, detailed information on their composition dependence is sparse. Here, we address this question and provide (Al,Ga)N band parameters for widely used multi-band $\mathbf{k}\cdot\mathbf{p}$ Hamiltonians. We start from density functional theory (DFT) to sample the electronic structure of (Al,Ga)N alloys over the full composition range. The $\mathbf{k}\cdot\mathbf{p}$ band parameters are treated as free parameters to reproduce the DFT data. For GaN and AlN the parameters extracted here agree well with literature values. When turning to the composition dependence of the $\mathbf{k}\cdot\mathbf{p}$ parameters, our calculations show that most parameters deviate significantly from a linear interpolation of the GaN and AlN values, an approximation widely made in the literature. Moreover, to describe changes in the band parameters with Al content, \emph{composition dependent} bowing parameters have to be considered for an accurate description of the DFT data. Finally, our analysis also provides initial insight into consequences of the nonlinear composition dependence of the $\mathbf{k}\cdot\mathbf{p}$ parameters for the electronic structure of (Al,Ga)N alloys. We find that in particular the band ordering is affected by the nonlinear evolution of the crystal field splitting energy with composition, which is an important aspect for the light polarization characteristics of high Al content (Al,Ga)N alloys.
\end{abstract}
\maketitle

\section{Introduction}

Optoelectronic devices operating in the ultraviolet (UV) spectral region are of interest for a variety of different applications, including biomedicine and disease diagnostics~\cite{westerhof1997treatment,panov2015application,dhankhar2020novel}. Furthermore, UV sterilization and 
disinfection~\cite{hirayama2017algan,clarke2006ultraviolet} received strong attention in fighting the coronavirus and its variants~\cite{houser2020ten}. Current UV light emission relies to a large extent on bulky, inefficient and wavelength non-tunable gas lasers~\cite{green1978new} or lamps utilizing toxic mercury~\cite{muramoto2014development}. In recent years, the semiconductor alloy aluminium gallium 
nitride, (Al,Ga)N, has been targeted for replacing these inefficient and environmental unfriendly devices as the direct (Al,Ga)N band gaps cover, in principle, the wavelength range from UV-A (315-400 nm) to UV-B (280-315) into the UV-C (200-280 nm)~\citep{kneissl2019emergence}.

Although (Al,Ga)N-based UV light emitters present an attractive option for solid-state-based solutions, the efficiency of these devices is currently low, especially in the deep UV (UV-C) spectral range~\cite{amano2020,lang2025progress}. Several aspects contribute to these low efficiencies, including low light extraction efficiencies (LEEs) and $p$-doping of (Al,Ga)N layers with high Al contents~\cite{kneissl2019emergence,amano2020}. In general, the low LEE  is tightly linked to fundamental differences in the valence band structure of AlN and GaN systems~\cite{liu2018influence}: The valence band edge (VBE) in AlN is of $\Gamma_{7}$ symmetry while in GaN it is of $\Gamma_9$ symmetry~\cite{suzuki1995first,coughlan2015band}. Optical transitions involving the conduction band edge (CBE) and VBE result in transverse magnetic (TM) polarized light in AlN while in GaN the emitted light is of transverse electric (TE) polarization~\cite{nam2004unique,banal2009optical}. Given that (Al,Ga)N based LEDs are mainly top or bottom emitting devices, TM polarized emission leads to a low LEE ~\cite{guttmann2019optical}. Moreover, in (Al,Ga)N QWs, which form the active region of an LED, several further factors impact the valence band ordering, including strain and carrier/quantum confinement, but also the magnitude and sign of the so-called crystal field splitting (CF) energy, $\Delta_\text{cf}$~\cite{liu2018influence}. For AlN, $\Delta_\text{cf}$ is negative (in the range $-160$ to $-295$ meV), while in GaN it is positive and significantly smaller in magnitude ($<40$ meV)~\cite{ponce2019hole, wei1996valence, vurgaftman2001band, vurgaftman2003band,rinke2008consistent}. Understanding how $\Delta_\text{cf}$ changes with alloy content plays an important role in designing (Al,Ga)N-based light emitters, as it will affect the band ordering and thus polarization characteristics of the emitted light. However, as mentioned above, not only is the crystal field splitting energy important but also quantum confinement effects will impact the valence band ordering in (Al,Ga)N heterostructures. Therefore, the effective electron and hole masses of the alloy along, for instance, the growth direction of a quantum well are key.

Device-level modeling of, for example, (Al,Ga)N-based LEDs largely relies on self-consistent Schr{\"o}dinger-Poisson solvers that are coupled with drift-diffusion simulations~\cite{chlipala2024harnessing,xing2022improvement}. For the wide-band gap system (Al, Ga)N, the Schr{\"o}dinger equation is widely treated in the framework of 6+2 band $\mathbf{k}\cdot \mathbf{p}$ models~\cite{fu2011exploring, gladysiewicz2019material, xing2024improving}. Such an approach requires, e.g. effective masses as well as crystal field splitting energies as input. Parameters for binary AlN and GaN systems have been extracted from experiment but more often from first-principles calculations. However, information on the composition dependence of these parameters in an (Al,Ga)N alloy is sparse. Often a linear interpolation of the binary parameters with the Al content $x$ is employed in an Al$_x$Ga$_{1-x}$N alloy~\cite{yamaguchi2010theoretical,wang2016enhancement,gladysiewicz2021material,shen2021three}. On the other hand, several studies have already indicated that, for example, the crystal field splitting energy exhibits deviations from such linear composition behavior ~\cite{neuschl2014composition, coughlan2015band}. Building on this, it is unclear whether a linear interpolation of effective mass parameters is justified.

In this work, we provide information on the composition dependence of {Al$_x$Ga$_{1-x}$N band parameters. More specifically, we report hole and electron effective mass parameters, which are required for 6+2 band $\mathbf{k}\cdot\mathbf{p}$ Hamiltonians that underlie widely available device simulation packages. We determine Luttinger-like $A_{i}(x)$ with $i=1,....,6$ parameters and effective electron mass $m_e(x)$ as a function of the AlN content $x$ over the entire range of composition. In addition, we also study the composition dependence of the crystal field splitting energy $\Delta_\text{cf}(x)$. To do so, we use density functional theory (DFT) to obtain effective electronic band structures of Al$_x$Ga$_{1-x}$N alloys which are then fitted with our 6+2 band $\mathbf{k}\cdot\mathbf{p}$ Hamiltonian. Our procedure relies on a Sobol sequence that recently has been used to obtain $\mathbf{k}\cdot\mathbf{p}$ parameters for Hamiltonians with larger basis sets, e.g. a 16 band Hamiltonian, from DFT band structures~\cite{marquardt2020multiband}. In general, we follow the procedure established by Rinke \emph{et al.}~\cite{rinke2008consistent} to extract $\mathbf{k}\cdot\mathbf{p}$ parameters. Our determined $\mathbf{k}\cdot\mathbf{p}$ parameters for AlN and GaN agree well with previously published values~\cite{vurgaftman2003band, rinke2008consistent}. Moreover, our calculations further support that the crystal-field splitting energy deviates noticeably from a linear interpolation of the binary material values. In addition, we find that the effective mass parameters deviate significantly from linearly interpolated values and can be strongly composition dependent. Finally, when using the extracted composition dependent $A_i(x)$ and $\Delta_\text{cf}(x)$ parameters for band structure calculations, we observe noticeable deviations in ordering and orbital character of the valence bands when compared to band structures predicted from linearly interpolated $A_i(x)$ and $\Delta_\text{cf}(x)$ values. The difference in band ordering can be of strong importance when investigating light polarization characteristics of (Al,Ga)N-based devices. Thus, our extracted parameters present an ideal starting point for device-level simulations to further understand the impact of the material parameters for light-emitting devices.       

The manuscript is organized as follows. We first describe the DFT approach underlying our calculations of (Al,Ga)N band structures in Sec.~\ref{sec:DFT}. The $\mathbf{k}\cdot\mathbf{p}$ model and the connected fitting procedure are discussed in Sec.~\ref{sec:kpfitting}. In Sec.~\ref{sec:results}, we present the results of our study, namely the composition dependent $m_e(x)$, $A_{i}(x)$, and $\Delta_\text{cf}(x)$ parameters for Al$_x$Ga$_{1-x}$N alloys. Moreover, we examine the impact of the obtained parameters on electronic bulk band structures of (Al,Ga)N alloys. 

\section{Theory}
 
In this section we outline the ingredients of the theoretical framework used to extract composition-dependent $\mathbf{k}\cdot\mathbf{p}$ parameters for (Al,Ga)N alloys. We start in Sec.~\ref{sec:DFT} with details of the DFT simulations, which form the basis of our $\mathbf{k}\cdot\mathbf{p}$ fitting procedure. In Sec.~\ref{sec:kpfitting}, details of the $\mathbf{k}\cdot\mathbf{p}$ model are given before turning to the fitting approach.  
 
\subsection{Density functional theory calculations}
\label{sec:DFT}

The DFT calculations carried out in this work are based on the plane-wave-based Vienna ab initio Simulation Package (VASP) version 5.4.4~\cite{hafner2008ab}.  We employ the generalized gradient approximation (GGA), as implemented in the Perdew–Burke–Ernzerhof (PBE) exchange-correlation functional, using the projector augmented wave (PAW) scheme. The semi-core $d$ electrons of Ga are treated as valence electrons and the plane wave cut-off energy is 600 eV.  The convergence criteria were $1 \times 10^{-4}$ eV for the energy minimisation and $1 \times 10^{-3}$ eV/\AA\, for the forces in the ionic relaxations. DFT calculations have been performed for unstrained bulk systems, thus we do not consider effects from an underlying substrate. Structural relaxation was achieved by minimizing the pressure on the supercell. As spin-orbit coupling (SOC) energies are small in nitrides when compared to other III-V materials (e.g. $\Delta^\text{GaN}_\text{so}= 17$ meV and $\Delta^\text{GaAs}_\text{so}=341$ meV) \cite{vurgaftman2001band, vurgaftman2003band},  we neglect SOC effects as it significantly increases the computational load while introducing only small changes to the band structure. This is a widely used approach in DFT calculations for III-N materials~\cite{yan2014effects, strak2017ab, goodrich2021prospects, ahmad2022polarization, liu2024chemical}.

The above outlined DFT framework has been chosen for several reasons for our $\mathbf{k}\cdot\mathbf{p}$ parameter extraction. Firstly, even though the PBE-GGA functional underestimates the band gap of GaN and AlN, previous literature studies show that effective \emph{electron} masses obtained from this DFT setup agree well with data extracted from functionals that predict band gaps closer to experiment~\citep{balestra2022electron,yan2014first}}. Secondly, in comparison to state-of-the-art  Heyd, Scuseria and Ernzerhof (HSE) hybrid functionals \citep{heyd2003hybrid,krukau2006influence}, the PBE-GGA functional is computationally less demanding. Therefore, it allows for a numerically efficient way to screen larger numbers of alloy configurations (see below) and Al contents $x$ to determine $\mathbf{k}\cdot\mathbf{p}$ parameters for Al$_x$Ga$_{1-x}$N alloys. Additionally, in case of HSE-DFT, the exact exchange mixing parameter, $\alpha$, has to be determined and specified. Previous studies have shown that to obtain a good description of the bulk band gaps of GaN and AlN, the required $\alpha$ values for the two materials, GaN ($\alpha$ = 0.294) and AlN ($\alpha$ = 0.337), are (very) different~\cite{moses2011hybrid}. Thus, for an alloy, there is uncertainty on how to determine $\alpha$. In contrast, in the PBE-GGA functional, there is no free parameter and therefore no further input is required for (Al,Ga)N alloys. Finally, and as we will show below, the valence band effective mass parameters extracted from our PBE-GGA calculations are in good agreement with literature values on GaN and AlN binary materials. Therefore, in terms of effective mass parameters, the PBE-GGA functional should provide a reliable framework over the full composition range in Al$_x$Ga$_{1-x}$N alloys.

For the binary GaN and AlN systems a 4 atom primitive unit cell with a $\Gamma$ centered  $6\times 6 \times 4$ $k$-point mesh has been used. To treat Al$_x$Ga$_{1-x}$N alloys we have employed a 72 atom supercell and considered Al contents of $x=0.11,0.25,0.50,0.75$ and $0.89$. In the supercell case, the $\Gamma$ centered $k$-point mesh is $3\times3\times2$. Starting from experimental findings ~\cite{coli2002excitonic, westmeyer2006determination, rigutti2018atom}, we treat (Al,Ga)N as a random alloy, thus Al and Ga atoms are randomly distributed in the supercell. To account for the impact of the alloy microstructure, for each alloy content $x$, the electronic structure of 10 random alloy configurations has been investigated.    

The discussed DFT framework can now provide electronic band structures for extracting $\mathbf{k}\cdot\mathbf{p}$ parameters. Details of the fitting procedure are given in the following section, where we first introduce the $\mathbf{k}\cdot\mathbf{p}$ model employed.

\subsection{$\mathbf{k}\cdot \mathbf{p}$ model and band structure fitting approach}
\label{sec:kpfitting}

Following the study by Rinke \emph{et al.}~\cite{rinke2008consistent}, our $\mathbf{k}\cdot \mathbf{p}$ Hamiltonian is expanded in the basis states \textbar\ensuremath{S}\textuparrow \textrangle,\textbar\ensuremath{X}\textuparrow \textrangle, \textbar\ensuremath{Y}\textuparrow \textrangle, \textbar\ensuremath{Z}\textuparrow \textrangle, \textbar\ensuremath{S}\textdownarrow\textrangle, \textbar\ensuremath{X}\textdownarrow\textrangle, \textbar\ensuremath{Y}\textdownarrow \textrangle, \textbar\ensuremath{Z}\textdownarrow\textrangle.   
The explicit form of the Hamiltonian is given in Appendix A.
Although in general the chosen basis states lead to a so-called 8-band model, thus accounting for the coupling between $s$-like conduction and $p_{x,y,z}$-like valence bands, device simulations of (Al,Ga)N heterostructures often treat valence and conduction bands separately~\cite{witzigmann2020calculation, hofmann2024simulation, guttmann2019effect}. This approximation is, in general, justified by the large band gap of the alloy ($>3.5$ eV). Decoupling conduction and valence bands helps reducing the computational burden when connecting self-consistent Schr{\"o}dinger-Poisson solvers with drift-diffusion models for device simulations. Bearing this in mind, we consider in the following a 2+6 band $\mathbf{k}\cdot \mathbf{p}$ model. However, it should be noted that the DFT band structures used for $\mathbf{k}\cdot\mathbf{p}$ parameter extraction account intrinsically for conduction and valence band coupling effects. Therefore, even though the fitting is carried out with a 2+6 band $\mathbf{k}\cdot\mathbf{p}$ model, coupling effects are implicitly built into the parameters obtained. 
 
When it comes to fitting $\mathbf{k}\cdot\mathbf{p}$ to DFT band structures, we treat the effective electron mass, $m_e$, valence band effective mass parameters, $A_i$ with $i=1,\dots, 6$, and crystal field splitting energy, $\Delta_\text{cf}$, as free parameters. Here, it should be noted that in an (Al,Ga)N alloy, disorder will also lead to band mixing effects~\cite{finn2022impact, finn2025theoretical}. On the one hand, it is important to capture these effects in the fitting procedure, as this can affect the polarization of the emitted light from (Al,Ga)N systems, and potentially the LEE of a device~\cite{finn2025theoretical}. On the other hand, no extra terms can be added to the $\mathbf{k}\cdot\mathbf{p}$ Hamiltonian discussed above and in Appendix A. This restriction is introduced to ensure that our model parameters are directly applicable to widely available (commercial) software packages, which, in general, only allow defining input parameters for the Hamiltonian but not a ``new'' Hamiltonian. To account for this constraint, but also to capture alloy disorder-induced band mixing effects and lifting of degeneracies, we proceed as follows. In general, SOC introduces band mixing effects but also a lifting of degeneracies in bulk $\mathbf{k}\cdot\mathbf{p}$ band structures. Therefore, we use this parameter to mimic the effect of alloy disorder on the band structure of (Al,Ga)N alloys. For example, at $\mathbf{k}=\mathbf{0}$ the energies of the three highest lying valence band states in a semiconductor with a wurtzite crystal structure can be written as follows~\cite{chuang1996k}:  

\begin{eqnarray}
\label{eq:VBE1}
E_{\text{VB}_1}&=&E_{v}+\Delta_\text{cf}+\frac{\Delta_\text{sp}}{3},\\
\nonumber
E_{\text{VB}_2}&=&E_{v}+\frac{3\Delta_\text{cf}-\Delta_\text{sp}}{6}\\
\label{eq:VBE2}
& &+\sqrt{\left(\frac{3\Delta_\text{cf}-\Delta_\text{sp}}{6}\right)^{2}+2\left(\frac{\Delta_\text{sp}}{3}\right)^{2}},\\
\nonumber
E_{\text{VB}_3}&=&E_{v}+\frac{3\Delta_\text{cf}-\Delta_\text{sp}}{6}\\
\label{eq:VBE3}
& &-\sqrt{\left(\frac{3\Delta_\text{cf}-\Delta_\text{sp}}{6}\right)^{2}+2\left(\frac{\Delta_\text{sp}}{3}\right)^{2}}\,\, .
\end{eqnarray}

Compared to Ref.~\cite{chuang1996k}, we assumed that $\Delta_\text{sp}=\Delta_{2}=\Delta_{3}$, which describes the SOC energy. Moreover, our $\Delta_\text{cf}$ corresponds to $\Delta_1$ in Ref.~\cite{chuang1996k} and is the crystal field splitting energy. In the absence of SOC and alloy disorder, two of the three valence bands (neglecting spin) are degenerate $E_{\text{VB}_1}=E_{\text{VB}_2}\neq E_{\text{VB}_3}$. In this case, the band-ordering depends on the sign of $\Delta_\text{cf}$. For our purpose, $\Delta_\text{sp}$ and $\Delta_\text{cf}$ are introduced as free parameters to account for alloy disorder induced valence band splitting effects in (Al,Ga)N alloys, especially at the $\Gamma$-point. Thus, in conjunction with DFT data on the energetic separation of the highest lying three valence bands at the $\Gamma$-point, Eqs.~(\ref{eq:VBE1})-(\ref{eq:VBE3}) are used to obtain an initial guess for $\Delta_\text{sp}$ and $\Delta_\text{cf}$ when targeting data away from the $\Gamma$-point.


In terms of the numerical routine used to fit $\mathbf{k}\cdot\mathbf{p}$ to DFT band structures and thus to extract the $\mathbf{k}\cdot\mathbf{p}$ parameters outlined above, we follow Marquardt~\emph{et al.}~\cite{marquardt2020multiband}. Here, we use a Monte Carlo fitting scheme~\cite{niederreiter1992random} based on a Sobol sequence method \citep{sobol1967distribution,niederreiter1988low}. Sobol sequences have been shown to yield superior results as compared to several other low-discrepancy point fitting techniques and as a natural substitute to usual random sequences; further details on the employed approach can be found in Refs.~\cite{marquardt2020multiband, marquardt2021simulating}. The $\mathbf{k}\cdot\mathbf{p}$ Hamiltonian and the Monte Carlo fitting scheme are implemented in the flexible, generalized multi-band $\mathbf{k}\cdot\mathbf{p}$ formalism within the plane-wave framework of the SPHInX library~\cite{marquardt2010plane,marquardt2014generalized,marquardt2021simulating}.

When it comes to selecting DFT data for $\mathbf{k}\cdot\mathbf{p}$ fitting, we follow largely the work by Rinke \emph{et al.}~\cite{rinke2008consistent}. Thus, for our $\mathbf{k}\cdot\mathbf{p}$ parameter extraction we chose the directions $\Sigma$, $\Lambda$, $T$, and $\Delta$ in the first Brillouin zone. More specifically, we select a) $\Gamma\rightarrow\frac{M}{20}$ ($\Sigma$ path), b) $\Gamma\rightarrow\frac{L}{20}$ ($\Lambda$ path), c) $\Gamma\rightarrow\frac{K}{20}$ ($T$ path) and d) $\Gamma\rightarrow\frac{A}{10}$ ($\Delta$ path). Given the underlying assumptions of the $\mathbf{k}\cdot\mathbf{p}$ method used here, i.e. perturbation theoretical approach at the $\Gamma$-point, our fitting procedure focuses on the region close to the $\Gamma$-point. For the $\mathbf{k}\cdot\mathbf{p}$ parameter extraction, equal path lengths have been covered. Furthermore, a high density of DFT k-points has been chosen along the different directions for accurate parameter extraction ($\geq$30 points). To determine the $\mathbf{k}\cdot\mathbf{p}$ parameters for AlN and GaN, the parameter sets of Vurgaftman and Meyer~\cite{vurgaftman2003band} have been used as an initial guess. To reduce the likelihood that the fitting procedure converges to a local minimum, the $\mathbf{k}\cdot\mathbf{p}$ parameters are allowed to be varied by plus minus the magnitude of the respective initial guess value. We have tested the impact of both changing initial guess values as well as the range over which the parameters are allowed to be varied on the resulting final parameter set. In general, both factors (initial guess and parameter range) had a negligible impact on the final $\mathbf{k}\cdot\mathbf{p}$ parameter set. For (Al,Ga)N alloys we generally employ the same procedure, except that the initial guess for our alloy $\mathbf{k}\cdot\mathbf{p}$ parameters is obtained as a linear interpolation of our GaN and AlN values. 

\begin{figure}[t!]
\includegraphics{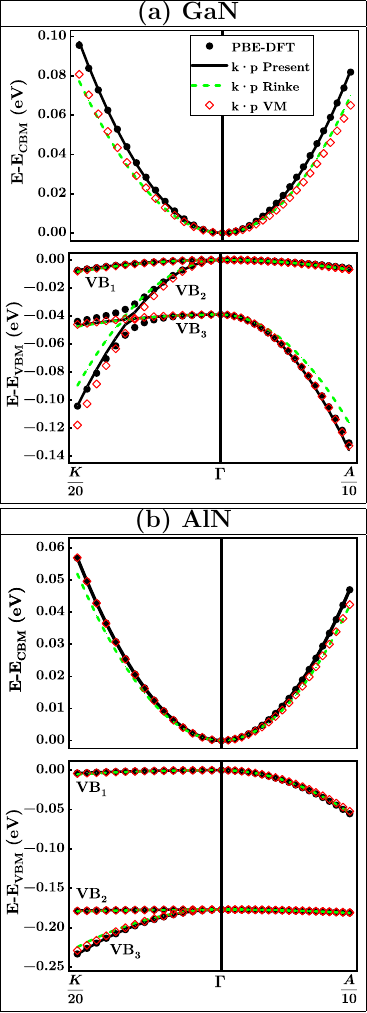}
\caption{Band structure of wurtzite (a) GaN and (b) AlN in the vicinity of the $\Gamma$ point. Our PBE DFT data are presented by filled black circles and the fitted $\mathbf{k}\cdot\mathbf{p}$ band structures by the solid black line. Band structures using the $\mathbf{k}\cdot\mathbf{p}$ parameter sets from Rinke \emph{et al.}~\cite{rinke2008consistent} and Vurgaftman and Meyer (VM) \cite{vurgaftman2003band} are given by dashed green lines and red open diamonds, respectively. To focus on differences in the dispersion, the band structures are plotted with respect to the conduction and valence band edge, respectively. Moreover, the crystal field splitting energy of our PBE-DFT calculations are used for all $\mathbf{k}\cdot\mathbf{p}$ calculations.}
\label{fig:fig1}
\end{figure}

\begin{table*}
\noindent
\centering{}\caption{Band parameters for AlN, GaN and (Al,Ga)N alloys when accounting for bowing parameters $b$, $\alpha$ and $\beta$. Definitions of these parameters are given in the main text.
Effective electron mass values parallel and perpendicular to the $c$-axis have been denoted by superscripts  $\Vert$ and $\bot$, respectively. Literature values are from Rinke \emph{et al.}~\cite{rinke2008consistent} and Vurgaftman and Meyer~\cite{vurgaftman2003band} are given for comparison. Percentage differences between values obtained in the present study and literature data are given in brackets.} 
\begin{tabular}{|c|c|c|c|c|c|c|c|c|c|}
\hline 
\multirow{3}{*} {\thead{Band\\Parameters}} & \multicolumn{6}{c|}{{Materials}} & \multirow{3}{*}{$b$} & \multirow{3}{*}{$\alpha$} & \multirow{3}{*}{$\beta$} \tabularnewline
\cline{2-2}  \cline{3-3} \cline{4-4} \cline{5-5} \cline{6-6} \cline{7-7}
& \multicolumn{3}{c|}{{GaN}} & \multicolumn{3}{c|}{{AlN}} & & &
\tabularnewline 
\cline{2-2}  \cline{3-3} \cline{4-4} \cline{5-5} \cline{6-6} \cline{7-7}
& Present & Ref.~\cite{rinke2008consistent} & Ref.~\cite{vurgaftman2003band} &  Present & Ref.~\cite{rinke2008consistent} & Ref.~\cite{vurgaftman2003band} & & &   \tabularnewline
\cline{2-2}  \cline{3-3} \cline{4-4} \cline{5-5} \cline{6-6} \cline{7-7} 
\hline $A_{1}$ & -7.438 & -5.947 (20.0\%) & -7.21 (3.1\%) & -4.101 & -3.991 (2.7\%) & -3.86 (5.9\%)& - & -5.48 & 3.46\tabularnewline
\hline $A_{2}$ & -0.485 & -0.528 (8.9\%) & -0.44 (9.3\%) & -0.211 & -0.311 (47.4\%) & -0.25 (18.5\%) & - & -1.07 & 6.35\tabularnewline
\hline $A_{3}$ & 6.965  & 5.414 (22.3\%)  & 6.68 (4.1\%)  & 3.822  & 3.671 (4.0\%)  &  3.58 (6.3\%) & - &  5.46 & 5.00 \tabularnewline
\hline $A_{4}$ & -2.963 & -2.512 (15.2\%) & -3.46 (16.8\%) & -1.487 & -1.147 (22.9\%) & -1.32 (11.2\%) & - & -0.29 & -0.89\tabularnewline
\hline $A_{5}$ & -2.998 & -2.510 (16.3\%)& -3.40 (13.4\%) & -1.632 & -1.329 (18.6\%) & -1.47 (9.9\%) & - & -2.40 & 0.33\tabularnewline
\hline $A_{6}$ & -4.352 & -3.202 (26.4\%) & -4.90 (12.6\%) & -2.11  & -1.952 (7.5\%) & -1.64 (22.3\%)& - & -3.84 & 0.88\tabularnewline
\hline $\Delta_\text{cf}$ (eV) & 0.040 & 0.034 (15.0\%)  & 0.010 (75.0\%)    & -0.177  & -0.295 (66.7\%) &-0.169 (4.52\%) & - & -0.19 (eV) & -0.22\tabularnewline
\hline $\Delta_\text{sp}$ (eV) &  -    & -      & -     & -      & -      & -     & - & -0.07 (eV) & -0.91\tabularnewline
\hline $m_e^{\bot}$ $(m_0)$ & 0.166 &  0.209(25.9\%) & 0.20 (20.0\%)& 0.30 & 0.329 (9.7\%) & 0.30 (0.0\%) & -0.060 & - & - \tabularnewline
\hline $m_e^{\Vert}$ $(m_0)$ & 0.150 & 0.186 (24.0\%) & 0.20 (33.3\%) & 0.287 & 0.322 (12.2\%) & 0.32 (11.5\%) & -0.052 & - & - \tabularnewline
\hline 
\end{tabular}
\label{table:table1}
\end{table*}

\section{Results}
\label{sec:results}

In this section we present (i) results of the above outlined parameter extraction procedure and (ii) consequences of the obtained composition dependent $\mathbf{k}\cdot\mathbf{p}$ parameters for the valence band structure of (Al,Ga)N alloys. In Sec.~\ref{sec:binary} we summarise our $\mathbf{k}\cdot\mathbf{p}$ parameters for GaN and AlN systems and compare the obtained data with available literature values.  The $\mathbf{k}\cdot\mathbf{p}$ parameters extracted for (Al,Ga)N alloys are discussed in Sec.~\ref{sec:AlGaN_kp_parameters}. Finally, in Sec.~\ref{sec:Comparison_BS} we compare electronic band structures of (Al,Ga)N alloys calculated using our composition dependent $\mathbf{k}\cdot\mathbf{p}$ parameters with band structures obtained when employing a linear interpolation of the values for GaN and AlN found in Sec.~\ref{sec:binary}.  

\subsection{GaN and AlN parameters}
\label{sec:binary}

Figure~\ref{fig:fig1} displays the DFT band structure together with the fitted $\mathbf{k}\cdot \mathbf{p}$ band structure for two of the four considered directions in the first Brillouin zone for (a) GaN and (b) AlN. The very good agreement between $\mathbf{k}\cdot\mathbf{p}$ and the DFT band structure highlights that both GaN and AlN are accurately described by our $\mathbf{k}\cdot\mathbf{p}$ model in the chosen $\mathbf{k}$ range. The extracted valence band $\mathbf{k}\cdot\mathbf{p}$ parameters $A_i$, electron effective mass values parallel, $m_e^{\Vert}$, and perpendicular, $m_e^{\bot}$, to the $c$-axis, as well as the crystal field splitting energy $\Delta_\text{cf}$ are summarized in Table~\ref{table:table1}. To compare our data with existing literature data, Fig.~\ref{fig:fig1} also displays $\mathbf{k}\cdot\mathbf{p}$ band structures when using the $A_i$ and $m_e$ parameter sets from Vurgaftman and Meyer~\cite{vurgaftman2003band} as well as Rinke \emph{et al.}~\cite{rinke2008consistent}. To focus on the band dispersion, the band structures are always plotted with respect to CBE and VBE. Moreover, the crystal field splitting energies from our calculations have been used for all parameter sets. Our DFT calculations give $\Delta^\text{AlN}_\text{cf}=-177$ meV for AlN  and $\Delta^\text{GaN}_\text{cf}=40$ meV  for GaN. The obtained AlN value is in good agreement with the value reported by Vurgaftman and Meyer \cite{vurgaftman2003band} ($\Delta^\text{AlN,VM}_\text{cf}=-169$ meV) but magnitude-wise it is noticeably smaller when compared to the data reported by Rinke \emph{et al.} ($\Delta^\text{AlN,Rinke}_\text{cf}=-295$ meV)~\cite{rinke2008consistent}. However, for GaN, our data on the crystal splitting energy agree well with the value reported by Rinke \emph{et al.} ($\Delta^\text{GaN,Rinke}_\text{cf}=34$ meV)~\cite{rinke2008consistent}. Vurgaftman and Meyer~\cite{vurgaftman2003band} suggest a smaller value ($\Delta^\text{GaN,VM}_\text{cf}=10$ meV). In general, one should note that the work in Ref.~\cite{vurgaftman2003band} is a summary and average of parameters reported in the literature.

Looking at the energy dispersion of AlN, Fig.~\ref{fig:fig1} (b) we find that over the $\mathbf{k}$-point range considered, all three parameter sets give band structures in very good agreement with each other. The different $\mathbf{k}\cdot \mathbf{p}$ parameter sets are summarized and compared in Table~\ref{table:table1}. For GaN, Fig.~\ref{fig:fig1} (a), we observe some deviations between three the $\mathbf{k}\cdot \mathbf{p}$ parameter sets, however, very close to the $\Gamma$-point the band structures do not differ significantly. In general, numbers may also vary slightly depending on the lengths of the path chosen for the fitting procedure.  Given the general good agreement between our data and the literature values for GaN and AlN, in the next section we extract $\mathbf{k}\cdot\mathbf{p}$ parameters for (Al,Ga)N alloys.

\begin{figure}
\includegraphics{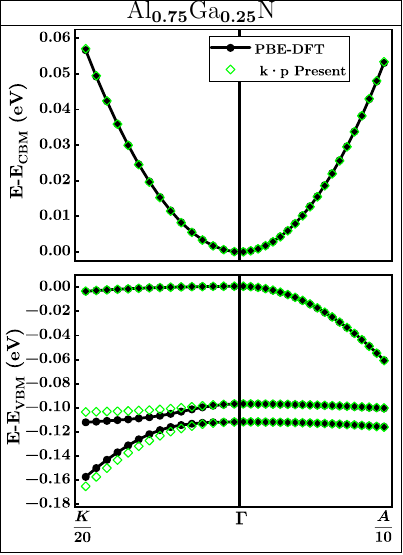} 
\caption{(a) Conduction and (b) valence band structure of an Al$_{0.75}$Ga$_{0.25}$N random alloy in the vicinity of the $\Gamma$ point. The black dotted lines denote our PBE-DFT results while the open green diamonds show the fitted $\mathbf{k}\cdot\mathbf{p}$ band structures.}
\label{fig:fig2}
\end{figure}

\begin{figure}[b]
\includegraphics{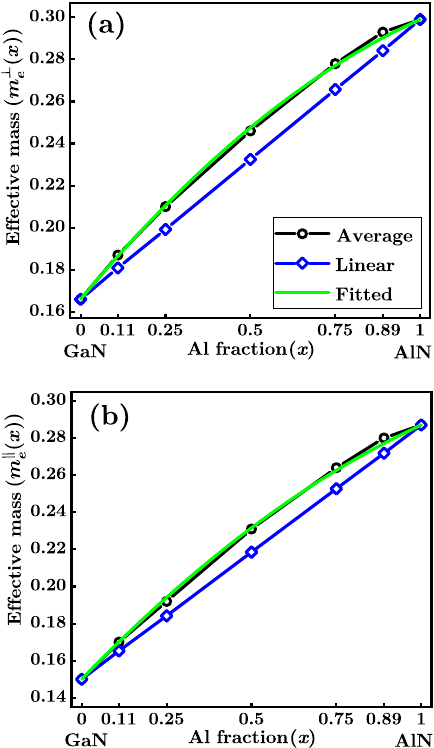}\\ 
\caption{Conduction band effective masses of (Al,Ga)N alloys for (a) perpendicular, $m_{e}^{\bot}$, 
and (b) parallel, $m_{e}^{\Vert}$, to the wurtzite $c$-axis.}
\label{fig:fig3}
\end{figure}

\begin{figure*}
\includegraphics{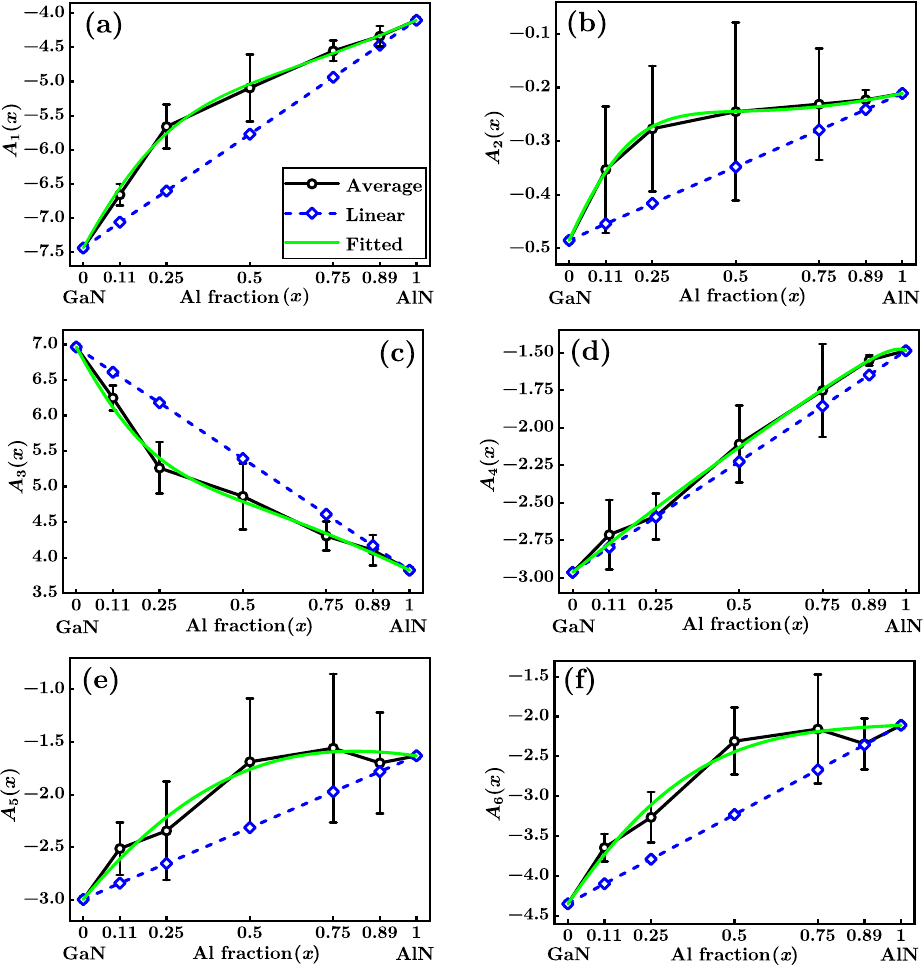}
\caption{Luttinger-like parameter $A_{i} (x)$ with $i=1,\ldots,6$ as a function of the alloy content $x$ of  Al$_x$Ga$_{1-x}$N alloys. The black circles represent the average value of $A_{i}$ parameters over 10 random alloy configurations per $x$; the blue diamonds denote the values obtained from a linear interpolation of the AlN and GaN $A_i$ values. The solid green lines are obtained using Eq.~(\ref{eq:eq4}); see main text for further details.}
\label{fig:fig5}
\end{figure*}

\begin{figure}[t!]
    \centering
    \includegraphics{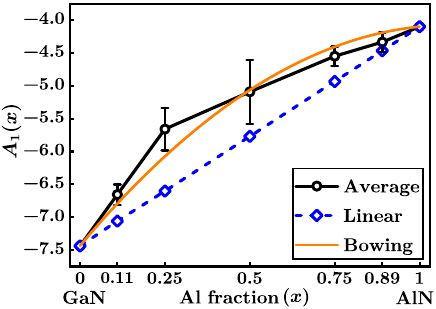}  
    \caption{Luttinger-like parameter $A_{1} (x)$ as a function of the Al content $x$ of  Al$_x$Ga$_{1-x}$N alloys. The black circles represent the average value of $A_{1}$ over 10 random alloy configurations per $x$ while the blue diamonds denote the values obtained from a linear interpolation of the AlN and GaN $A_1$ values. The solid orange line is obtained by fitting the data with a composition \emph{independent} bowing parameter, Eq.~\ref{eq:eq3}. See main text for further details.}
    \label{fig:fig4}
\end{figure}

\subsection{(Al,Ga)N alloy parameter}
\label{sec:AlGaN_kp_parameters}
In this section we focus on the composition dependence of the $\mathbf{k}\cdot\mathbf{p}$ parameters of Al$_x$Ga$_{1-x}$N alloys. We start by investigating the effective electron mass, both $m_e^{\Vert}(x)$ and $m_e^{\bot}(x)$, of (Al,Ga)N alloys. Subsequently, we study the valence band effective mass parameters $A_i(x)$. As discussed in Sec.~\ref{sec:DFT}, we us 10 different random alloy configurations per Al content $x$ in the considered Al$_{x}$Ga$_{1-x}$N alloys and fit each individual band structure with the $\mathbf{k}\cdot\mathbf{p}$ model. The final parameter set is then obtained as an average over the 10 different configurations. Figure~\ref{fig:fig2}  displays an example of our $\mathbf{k}\cdot\mathbf{p}$ fit to an effective Al$_{0.75}$Ga$_{0.25}$N DFT band structure (arbitrarily selected alloy configuration). As Fig.~\ref{fig:fig2} shows, the $\mathbf{k}\cdot\mathbf{p}$ band structure is in excellent agreement with the DFT data over the $\mathbf{k}$-point range considered. This highlights that (i) the $\mathbf{k}\cdot\mathbf{p}$ model is suitable for describing the alloy band structure and (ii) that the SOC-like term in our $\mathbf{k}\cdot\mathbf{p}$ Hamiltonian captures the alloy-induced lifting of degeneracies in the band structure well. 
We note that alloy disorder breaks the translational symmetry of the crystal, which results in the situation that Bloch's theorem is in principle no longer applicable. Thus, defining a band structure is not strictly valid. However, $\mathbf{k}\cdot\mathbf{p}$ descriptions of alloys assume that perturbations of disorder are weak and that an effective band structure can still be assumed. To bridge the gap between disorder and effective band structures, we have constructed supercells that are large enough to capture disorder effects on an atomistic level but small enough so that electrons with low $\mathbf{k}$-vectors move in the periodic potential due to the periodic boundary conditions of the plane-wave DFT approach used. In doing so, effects of alloy disorder are included in a limit that is consistent with the underlying assumptions of a $\mathbf{k}\cdot\mathbf{p}$ model.     

\subsubsection{Effective electron masses}

To gain insight into the composition dependence of the effective electron mass in (AlGa)N alloys, we follow here previous literature, and distinguish between effective electron masses perpendicular, $m_e^{\bot}(x)$, and parallel, $m_e^{\Vert}(x)$, to the $c$-axis. The effective electron masses are extracted according to the above outlined fitting procedure considering the in-plane directions $\Sigma$, $\Lambda$, $T$, and the out-of-plane direction $\Delta$ in the first Brillouin zone. The obtained masses are displayed in Fig.~\ref{fig:fig3} as a function of the Al content $x$ in Al$_{x}$Ga$_{1-x}$N alloys. In Fig.~\ref{fig:fig3} (a) the mass $m_{e}^{\bot}$ perpendicular to the $c$-axis is shown while (b) displays the mass $m_{e}^{\Vert}$ parallel to the $c$-axis. To evaluate the impact of alloy content $x$ on the effective electron mass, Fig.~\ref{fig:fig3} also shows the effective mass values for the two directions obtained by a linear interpolation of the effective mass values of GaN and AlN via:
\begin{equation}
m_{e,\text{AlGaN}}^{\alpha}=xm_{e,\text{AlN}}^{\alpha}+(1-x)m_{e,\text{GaN}}^{\alpha}\,\, ,
\label{eq:standardBowElec}
\end{equation}
where $\alpha=\{\bot,\Vert\}$ denotes the directions. 
The corresponding data are given in Fig.~\ref{fig:fig3} by the blue line (open blue diamonds). We note that different approaches have been used in the literature to estimate the effective electron mass of an alloy~\cite{krijn1991heterojunction, schoche2013electron, li2017localization}. However, to compare our results with recent literature data by Balestra~\emph{et al.}~\cite{balestra2022electron}, we use Eq.~(\ref{eq:standardBowElec}) as their DFT data indicate a Vegard law-type behavior for the composition dependence of the effective electron masses in (Al,Ga)N. From Fig.~\ref{fig:fig3} we infer that both for $m_{e}^{\bot}$ and $m_{e}^{\Vert}$ our data deviate from a linear interpolation of the effective electron masses via Eq.~(\ref{eq:standardBowElec}). It should be noted that our extracted electron masses do not differ significantly when extracted from the 10 different microscopic configurations at each Al content $x$ (the largest deviation is 1.65\% from the average value). The differences between our results and Ref.~\cite{balestra2022electron} may be due to several factors. Although in general the approach used in Ref.~\cite{balestra2022electron} is similar to ours, which means that it also utilizes PBE-DFT, the region over which the band structure is fitted to extract the effective electron masses may be different. Furthermore, the supercell employed in Ref.~\cite{balestra2022electron} is smaller (32-atoms) when compared to ours (72 atoms).Further studies would be required elucidate on the differences between our work and Ref.~\cite{balestra2022electron}.

To provide an improved description of the composition dependence of the effective electron masses but also to extrapolate to composition values not explicitly studied here, we adapt the widely used expression for the band gap evolution~\cite{koide1987energy, kuo2002band} of semiconductor alloys to effective electron masses and introduce a bowing parameter:
\begin{equation} \label{eq:eq2}
m_{e,\text{AlGaN}}^{\alpha}=xm_{e,\text{AlN}}^{\alpha}+(1-x)m_{e,\text{GaN}}^{\alpha}-b(1-x)x\,\, .
\end{equation}
Figure~\ref{fig:fig3} reveals that $m_{e}^{\bot}(x)$ and $m_{e}^{\Vert}(x)$ are indeed well described by Eq.~(\ref{eq:eq2}), thus when introducing the bowing parameter $b$. The effective electron mass bowing parameters obtained are summarized in Table~\ref{table:table1}.

\subsubsection{Valence band $\mathbf{k}\cdot \mathbf{p}$ parameters}

Having discussed the composition dependence of the electron effective mass in (Al,Ga)N alloys, we turn now and focus on the composition dependence of the valence band effective mass parameters $A_{i}$ with $i=1....6$, crystal field splitting energy, $\Delta_\text{cf}$, and the alloy disorder parameter, $\Delta_\text{sp}$. We start the analysis with the parameters $A_{i}(x)$, which are shown in Fig.~\ref{fig:fig5} as a function of the Al content $x$ in Al$_x$Ga$_{1-x}$N alloys. The black open circles give the average values of $A_{i}(x)$, including error bars, from the fitting to PBE-DFT band structures. Again, to visualize the impact of alloy disorder on $A_{i}(x)$, Fig.~\ref{fig:fig5} also shows $A_{i}(x)$ parameters obtained (blue open diamonds) when employing the often used linear interpolation (Vegard's law) of the GaN and AlN parameters with composition $x$~\citep{yamaguchi2010theoretical,wang2016enhancement,gladysiewicz2021material,shen2021three}.
Except for $A_{4}(x)$, the data obtained from a linear interpolation noticeably deviates from the PBE-DFT values.

Moreover, when trying to describe the composition dependence of the $A_i(x)$ parameters similar to the effective electron mass values via a composition \emph{independent} bowing parameter:
\begin{equation} \label{eq:eq3}
A_{i}^\text{AlGaN}=xA_{i}^\text{AlN}+(1-x)A_{i}^\text{GaN}-b^\text{A$_i$}_i(1-x)x \,\, ,
\end{equation}
the PBE-DFT is poorly approximated. An example for such a poor fitting to the PBE-DFT data via Eq.~(\ref{eq:eq3}) is given in Fig.~\ref{fig:fig4} for $A_1(x)$ as an example.  A similar behavior is observed for the other $A_{i}(x)$ parameters (not shown here). In general, this finding suggests that the bowing parameter itself is composition \emph{dependent}. A similar behavior has been observed for the band gap bowing in (In,Ga)N~\cite{van1999large, caro2013theory} but in particular in (Al,In)N alloys~\cite{sakalauskas2010dielectric,schulz2013composition, alam2020bandgap}.  In case of the band gap bowing of (Al,In)N, Sakalauskas \emph{et al.}~\citep{sakalauskas2010dielectric} introduced an empirical formulae to account for the composition dependence of the bowing parameter. We have adopted this approach here for the valence effective mass parameters $A_i(x)$:
 \begin{eqnarray} 
\nonumber
A_{i}^\text{AlGaN} &=& xA_{i}^\text{AlN}+(1-x)A_{i}^\text{GaN}-\frac{\alpha_i}{1+\beta_i x^{2}}(1-x)x\,\, \\
&=& xA_{i}^\text{AlN}+(1-x)A_{i}^\text{GaN}-\tilde{b}^{\text{A$_i$}}_i(x)(1-x)x\,\, .
\label{eq:eq4}
\end{eqnarray}
Here, $\alpha_i$ and $\beta_i$ are free parameters which are used to fit the averaged PBE-DFT data for each $A_i(x)$. Figure~\ref{fig:fig5} shows indeed that Eq.~(\ref{eq:eq4}) allows to capture the composition dependence of the $A_i$ parameters over the full composition range. Moreover, Eq.~(\ref{eq:eq4}) enables us also to predict $A_i$ parameters at compositions not explicitly targeted in the PBE-DFT calculations. The obtained $\alpha_i$ and $\beta_i$ values for the different $A_i$ parameters are summarized in Table~\ref{table:table1}.
\begin{figure*}
    \includegraphics{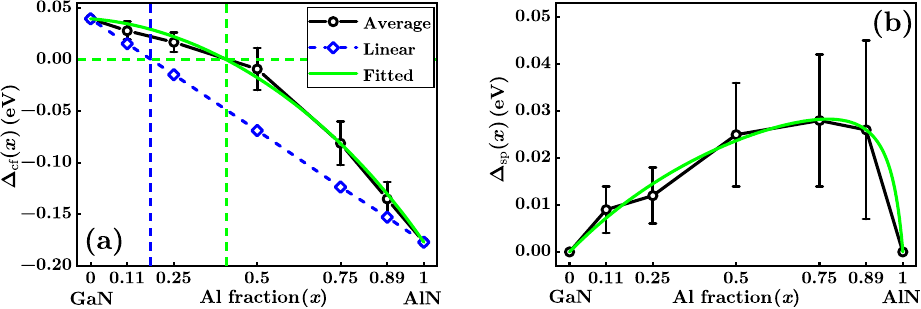}  
    \caption{Composition dependence of (a) crystal field splitting energy $\Delta_\text{cf}(x)$ and (b) the alloy disorder parameter $\Delta_\text{sp}(x)$ in (Al,Ga)N alloys.}
    \label{fig:fig6}
\end{figure*}

Having presented the results for the effective mass parameters $A_{i}(x)$, we turn in a second step to the composition dependence of the crystal field splitting energy, $\Delta_\text{cf}$, and the alloy disorder parameter, $\Delta_\text{sp}$. Figure~\ref{fig:fig6} (a) shows $\Delta_\text{cf}(x)$ as a function of the Al content $x$ in Al$_x$Ga$_{1-x}$N while Fig.~\ref{fig:fig6} (b) depicts the data for $\Delta_\text{sp}(x)$. Looking at Fig.~\ref{fig:fig6} (a), one observes that the $\Delta_\text{cf}(x)$ values extracted from PBE-DFT (black open circles) noticeably deviate from a linear interpolation (blue open diamonds) of the GaN and AlN crystal field splitting values. The evolution of $\Delta_\text{cf}$ with $x$ is again well described by a \emph{composition-dependent} bowing parameter of the form $\tilde{b}_\text{cf}(x)=\alpha_\text{cf}/(1+\beta_\text{cf} x^2)$, as used for the $A_i$ parameters in Eq.~(\ref{eq:eq4}). The extracted values $\alpha_\text{cf}$ and $\beta_\text{cf}$ are given in Table~\ref{table:table1}. As discussed above, the sign of $\Delta_\text{cf}$ plays an important role for the polarization characteristics of light emitted from (Al,Ga)N alloys. Assuming that carriers only occupy the highest valence band and in the absence of band mixing effects, a positive (negative) crystal field splitting energy $\Delta_\text{cf}(x)$ suggests TE (TM) polarized light emission. Given the non-linear behavior of $\Delta_\text{cf}(x)$ with composition $x$, Fig.~\ref{fig:fig6} (a) indicates that the onset of a polarization switching from TE to TM occurs at higher Al contents $x$ when compared to a linear interpolation of $\Delta_\text{cf}(x)$ with composition $x$. The dashed lines show the critical Al content $x_\text{cr}$ at which the sign of $\Delta_\text{cf}$ changes. In case of a linear interpolation of $\Delta_\text{cf}$, $x^\text{lin}_\text{cr}\approx0.18$. When accounting for the composition-dependent bowing parameter $\tilde{b}_\text{cf}(x)$ for predicting $\Delta_\text{cf}(x)$, $x^\text{bow}_\text{cr}\approx0.41$.

In general, previous studies also come to the conclusion that $\Delta_\text{cf}(x)$ deviates from the linear interpolation of the binary values of GaN and AlN~\cite{coughlan2015band,neuschl2014composition}. However, a direct comparison with existing literature data is difficult. Firstly, experimental studies providing information on the polarization switching are usually carried out on samples that are strained to a substrate. One can in principle account for strain effects when estimating the crossover for the unstrained case, although, it requires information on (i) the valence band deformation potentials and their composition dependence and (ii) crystal field splitting energies of the binary materials. As already discussed in Sec.~\ref{sec:binary}, in the literature there is a large spread in AlN and GaN values for $\Delta_\text{cf}$ and the valence and conduction band deformation potentials~\cite{yan2009strain, sheerin2022strain}. The closest to our study here are the empirical tight-binding model (ETBM) calculations by Coughlan \emph{et al.}~\cite{coughlan2015band}, which predict that $\Delta_\text{cf}$ changes sign at $x^\text{ETBM}_\text{cr}=0.25$. However, the calculations in Ref.~\cite{coughlan2015band} use different binary $\Delta_\text{cf}$ values for GaN, $\Delta^\text{GaN, ETBM}_\text{cf}=23$ meV, and AlN, $\Delta^\text{AlN, ETBM}_\text{cf}=-240$ meV, when compared to the values obtained here from PBE-DFT calculations ($\Delta^\text{GaN}_\text{cf}=40$ meV and $\Delta^\text{AlN}_\text{cf}=-177$ meV). Using our composition-dependent bowing parameter formulae with $\Delta^\text{GaN}_\text{cf}=23$ meV and $\Delta^\text{AlN}_\text{cf}=-240$ meV from Ref.~\cite{coughlan2015band}, we obtain that $\Delta_\text{cf}$ changes sign at $x_\text{cr}=0.20$, which is close to the value obtained by Coughlan \emph{et al.}~\cite{coughlan2015band}. Thus, the here obtained composition-dependent bowing parameter reflects well the ETBM results in Ref.~\cite{coughlan2015band} when accounting for differences in the crystal field splitting energies.

As already mentioned above, the sign of the crystal field splitting energy gives a first indication for changes in the valence band structure. However, alloy disorder also leads to band mixing effects and a lifting of degeneracies. We mimicked this effect through an SOC-like interaction term, introducing a single free parameter, $\Delta_\text{sp}$, in the Hamiltonian. In doing so, the developed method is compatible with $\mathbf{k}\cdot\mathbf{p}$ Hamiltonians underlying widely available (commercial) software packages. Figure~\ref{fig:fig6} (b) displays $\Delta_\text{sp}$ as a function of the Al content $x$. The data is again fitted with a bowing parameter of the form $\tilde{b}_\text{cf}(x)=\alpha_\text{cf}/(1+\beta_\text{cf} x^2)$; $\alpha_\text{cf}$ and $\beta_\text{cf}$ values for $\Delta_\text{sp}$ are summarized in Table~\ref{table:table1}. As there is no alloy disorder in GaN and AlN, for the binary systems $\Delta_\text{sp}=0$ meV. Overall, we observe that $\Delta_\text{sp}$ can be of similar magnitude to $\Delta_\text{cf}$ in the Al content composition range up to 50-60\%. For higher Al contents $\Delta_\text{sp}< |\Delta_\text{cf}|$. We also observe that $\Delta_\text{sp}$ is largest between $x=0.75$ and $x=0.89$, thus indicating that alloy disorder impacts the electronic structure in the high AlN content regime more strongly. A similar observation has been made by Finn \emph{et al.} when studying the electronic structure of (Al,Ga)N QWs by means of an atomistic ETBM~\cite{finn2022impact}.

\subsection{Impact of composition dependent $\mathbf{k}\cdot\mathbf{p}$ parameter on (Al,Ga)N band structures}
\label{sec:Comparison_BS}

As discussed above, $\Delta_\text{cf}$ and $\Delta_\text{sp}$ will impact both valence band ordering and mixing, which will affect the light polarization characteristics of (Al,Ga)N alloys. In the following, we compare $\mathbf{k}\cdot\mathbf{p}$ band structures of (Al,Ga)N alloys where the underlying parameters are determined by using (i) the composition-dependent bowing parameters, $\tilde{b}$, obtained in the present work and (ii) when linearly interpolating the AlN and GaN parameters and thus also neglecting alloy disorder-induced band mixing effects. Our aim is a general study of changes in the band structure rather than providing detailed insight into light polarization characteristics as this requires accounting for carrier densities and temperature effects, e.g., via a Fermi-Dirac distribution of carriers in the different valence bands. However, we can provide initial information on how the $\mathbf{k}\cdot\mathbf{p}$ parameters can affect the light polarization characteristics of (Al,Ga)N alloys by studying not only the energy dispersion of the different valence bands but also by resolving the orbital character of the different bands as a function of $\mathbf{k}$. For this analysis, we have selected two alloy compositions, namely Al$_{0.25}$Ga$_{0.75}$N and Al$_{0.50}$Ga$_{0.50}$N.  

\begin{figure*}[t!]
\includegraphics{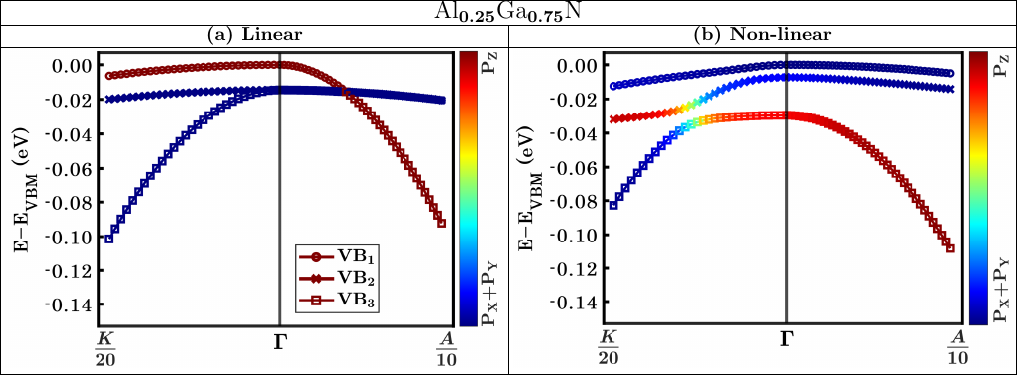}
\caption{Band structure of an Al$_{0.25}$Ga$_{0.75}$N alloy obtained from (a) linearly interpolated $\mathbf{k}\cdot\mathbf{p}$ parameters and (b) when accounting for alloy disorder effects, $\Delta_\text{sp}$, and composition dependent bowing parameters, $\tilde{b}(x)$.}
\label{fig:fig7}
\end{figure*}

In a first step, we study the $\mathbf{k}\cdot\mathbf{p}$ band structure of an Al$_{0.25}$Ga$_{0.75}$N alloy. For this Al content, $\Delta_\text{cf}<0$ when using a linear interpolation of the GaN and AlN values, as shown in Fig.~\ref{fig:fig6} (a). However, when accounting for the composition-dependent bowing parameter, $\tilde{b}(x)$, $\Delta_\text{cf}>0$. As mentioned above, the sign of $\Delta_\text{cf}$ gives first insight into the valence band ordering but does not provide information about band mixing effects or energetic separation of the different valence bands due to disorder.  Figure~\ref{fig:fig7} (a) shows the band structure of an Al$_{0.25}$Ga$_{0.75}$N alloy in the vicinity of the $\Gamma$-point for a linear interpolation of the $\mathbf{k}\cdot\mathbf{p}$ parameters with composition, while Fig.~\ref{fig:fig7} (b) depicts the band structure when accounting for composition-dependent bowing parameters, $\tilde{b}(x)$. The color coding of the bands gives the orbital character. We have grouped $p_x$ and $p_y$-like orbital contributions together as they both facilitate TE polarized light emission in contrast to $p_z$-like states which contribute to, in general, TM polarization. Looking at the band structure obtained from the linearly interpolated $\mathbf{k}\cdot\mathbf{p}$ parameters, Fig.~\ref{fig:fig7} (a), one finds that the topmost valence band, denoted as VB$_1$ in the following, is entirely $p_z$-like in character. The lower lying valence states, denoted as VB$_2$ and VB$_3$, are $p_x$ and $p_y$-like in character. The valence bands $\text{VB}_2$ and $\text{VB}_3$ are degenerate here since neither alloy disorder-related nor small SOC effects are considered in the linearly interpolated $\mathbf{k}\cdot\mathbf{p}$ parameter set. Overall, the band ordering seen in Fig.~\ref{fig:fig7} (a) is consistent with a negative $\Delta_\text{cf}$ value (cf. Fig.~\ref{fig:fig6} (a)), which means that the topmost valence band is expected to be $p_z$-like in character. However, even though VB$_1$ is $p_z$-like in character and thus suggesting TM polarized light emission, VB$_2$ and VB$_3$ are only approximately 20 meV below VB$_1$. Given that the thermal energy at room temperature ($T=300$ K) is around 24 meV, one can expect some occupation of VB$_2$ and VB$_3$ with holes at room temperature. This in turn will impact the degree of optical polarization, $\rho$, which can be defined as the ratio of intensities of TE to TM polarized light emission, $\rho = (I_\text{TE}- I_\text{TM})/(I_\text{TE}+ I_\text{TM}$)~\cite{coughlan2015band,li2022effect, lu2022enhancing} and potentially the LEE in a device. However, in general, the band structure depicted in Fig.~\ref{fig:fig7} (a) is indicative of predominantly TM polarized light emission in an Al$_{0.25}$Ga$_{0.75}$N alloy when using a linear interpolation of the $\mathbf{k}\cdot\mathbf{p}$ parameters with Al content.  

When accounting for the extracted composition-dependent bowing parameters, Fig.~\ref{fig:fig7} (b) reveals that the resulting $\mathbf{k}\cdot\mathbf{p}$ valence band structure noticeably deviates from the band structure obtained within the linear interpolation scheme, cf. Fig.~\ref{fig:fig7} (a). Firstly, the topmost valence band, VB$_1$, is mainly $p_x$+$p_y$-like in character at the $\Gamma$ point when accounting for composition dependent bowing parameters. Also VB$_2$ is predominantly of $p_x$+$p_y$-like character near $\Gamma$.  The degeneracy of VB$_1$ and VB$_2$ is lifted due to alloy disorder, here introduced by the $\Delta_\text{sp}$ term in the $\mathbf{k}\cdot\mathbf{p}$ Hamiltonian. The energetically lowest valence band, VB$_3$, is of $p_z$-like character. It is to note that VB$_3$ lies more than 30 meV below the valence band edge. Therefore, thermal occupation of this band at room temperature is unlikely, at least at low carrier densities.  Even though there is some band mixing between VB$_2$ and VB$_3$ along the $\Gamma\rightarrow K$ direction away from $\Gamma$, predominantly TE polarized light emission is expected when accounting for the composition-dependent bowing parameters, even at elevated temperatures. 
\begin{figure*}[t!]
\includegraphics{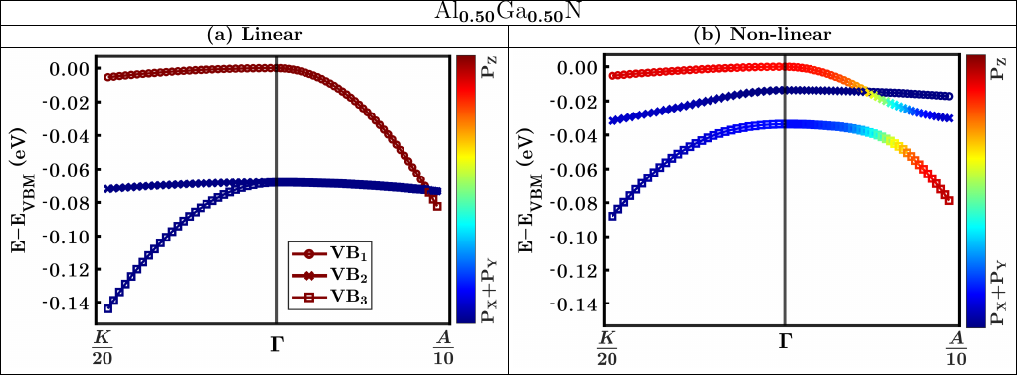}
\caption{Band structure of an Al$_{0.50}$Ga$_{0.50}$N alloy obtained from (a) linearly interpolated $\mathbf{k}\cdot\mathbf{p}$ parameters and (b) when accounting for alloy disorder effects, $\Delta_\text{sp}$, and composition dependent bowing parameters, $\tilde{b}(x)$.}
\label{fig:fig8}
\end{figure*}

We note again that our discussion is based on unstrained bulk systems. Strain or quantum confinements in, e.g., QWs can lead to further changes in the band structure and band-ordering as discussed in the literature~\cite{suzuki1996strain,finn2025theoretical,shen2022analysis}.
The effective masses of the different bands play an important role in understanding the impact of quantum confinement on, for instance, the valence band ordering in a QW system and thus the optical properties. In general, UV light emitters that utilize (Al,Ga)N alloys are grown along the wurtzite $c$-axis~\cite{sohi2017critical, tyagia2019algan, chen2022hole}. Therefore, the effective masses associated with the different valence bands toward the $A$ point (along the wurtzite $c$-axis) are important and we can now gain information on how these masses depend on the interpolation of the $\mathbf{k}\cdot\mathbf{p}$ parameters. For the calculated band structure, we employed both the analytic expressions provided in Appendix B, and numerical differentiation of the band dispersion in the vicinity of the 
$\Gamma$ point to extract the effective masses. Both approaches yield consistent results.  Figure~\ref{fig:fig7} (a) reveals a low effective hole mass ($m_{\text{VB}_1}=0.15$  m$_0$) in this direction for the highest valence band, VB$_1$, when using linearly interpolated parameters for the band structure calculation. The energetically lower-lying bands, VB$_2$ and VB$_3$, exhibit much higher effective masses ($m_{\text{VB}_2}=2.35$ m$_0$; $m_{\text{VB}_3}=2.35$ m$_0$). This suggests that quantum confinement effects will significantly impact the valence band ordering, as the topmost valence (VB$_1$) will shift to lower energies more strongly when compared to VB$_2$ and VB$_3$. Given the differences in the orbital character of the different valence bands, strong band mixing effects may be expected in a $c$-plane Al$_{0.25}$Ga$_{0.75}$N QW depending on e.g. the well width. This will also affect the degree of optical polarization and potentially the LEE. However, when accounting for nonlinearities in the $\mathbf{k}\cdot\mathbf{p}$ parameter interpolation the situation is different, as Figure~\ref{fig:fig7} (b) shows. In this case, VB$_1$ and VB$_2$ exhibit heavier effective masses ($m_{\text{VB}_1}=2.78$ m$_0$, $m_{\text{VB}_2}=1.04$ m$_0$) toward the $A$ point, while VB$_3$ has a low effective mass ($m_{\text{VB}_3}=0.19 $ m$_0$) in this direction. This means that quantum confinement along the wurtzite $c$-axis will shift VB$_3$ away from the valence band edge formed by VB$_1$ and VB$_2$. Given that VB$_1$ and VB$_2$ are predominately $p_x$ and $p_y$-like in orbital character, the energetically highest valence states in an Al$_{0.25}$Ga$_{0.75}$N QW will be formed by such orbitals, indicating mainly TE polarized light emission. This indicates a very different result compared to the conclusions drawn from the linearly interpolated $\mathbf{k}\cdot\mathbf{p}$ parameter set presented in Fig.~\ref{fig:fig7} (a).

In a second step, we turn our attention to the band structure of Al$_{0.50}$Ga$_{0.50}$N alloys, which often form the active region or the cladding layer of UV light emitters in the 250-300 nm wavelength range~\cite{zhang2017hole, asif2013substrate, chu2020impact, khan2023milliwatt}. Figure~\ref{fig:fig8} (a) shows the band structure resulting from a linear interpolation of the $\mathbf{k}\cdot\mathbf{p}$ parameters with Al content $x=0.5$. The band structure is largely identical to the band structure displayed in Figure~\ref{fig:fig7} (a) for the Al$_{0.25}$Ga$_{0.75}$N case: The energetically highest valence band, VB$_1$, is of $p_z$-like character, while the energetically lower-lying valence bands, VB$_2$ and VB$_3$, are $p_x+p_y$-like in character and degenerate at the $\Gamma$-point. The main difference to the  Al$_{0.25}$Ga$_{0.75}$N case is that the energetic separation between VB$_1$ and VB$_{2,3}$ is much greater at the $\Gamma$ point. In the Al$_{0.50}$Ga$_{0.50}$N alloy, the energetic separation is $>$ 60 meV, compared to only 20 meV for Al$_{0.25}$Ga$_{0.75}$N. This behavior is expected when looking at the linear interpolation of the crystal field splitting energy with composition (cf. Fig.~\ref{fig:fig6} (a)). As a consequence and in contrast to Al$_{0.25}$Ga$_{0.75}$N, thermal occupation of VB$_2$ and VB$_3$ is unlikely in the bulk case. Thus, the light emitted from a bulk Al$_{0.50}$Ga$_{0.50}$N alloy is expected to be TM polarized, at least when using linearly interpolated $\mathbf{k}\cdot\mathbf{p}$ parameters. Turning to the effective masses of the different valence bands toward the $A$-point, we observe a situation similar to Al$_{0.25}$Ga$_{0.75}$N, Fig.~\ref{fig:fig7} (a), in the sense that VB$_1$ ($m_{\text{VB}_1}=0.17$ m$_0$) exhibits a low effective mass while VB$_2$ and VB$_3$ ($m_{\text{VB}_2}= 2.66$ m$_0$, $m_{\text{VB}_3}=2.66$ m$_0$) have a high effective mass along this direction. Thus, quantum confinement in $c$-plane Al$_{0.50}$Ga$_{0.50}$N QWs can reduce the energetic separation between $p_z$ and $p_x$+$p_y$-like valence bands. However, given the large crystal field splitting energy, even in a strong carrier confinement regime, e.g. narrow wells, and when focusing on energetic shifts in the band structure resulting from the different effective masses of the valence bands only (no strain contributions), TM polarized emission may still be expected in the case of the linearly interpolated $\mathbf{k}\cdot\mathbf{p}$ parameters. Strain effects may change this behavior and would have to be further analyzed.   

The situation is again different when accounting for composition-dependent bowing parameters in determining the $\mathbf{k}\cdot\mathbf{p}$ parameters of the Al$_{0.50}$Ga$_{0.50}$N alloy, as the band structure in Fig.~\ref{fig:fig8} (b) clearly shows. Although the energetically highest valence band, VB$_1$, also has a larger $p_z$ orbital character (87\%) near the $\Gamma$ point, the energy separation to VB$_2$ is much smaller ($\approx 20$ meV) when compared to the band structure obtained from a linearly interpolated set of $\mathbf{k}\cdot\mathbf{p}$ parameters; see Fig.~\ref{fig:fig8} (a). Thus, at room temperature, occupation of VB$_2$ with holes can be expected. Moreover, the orbital character of VB$_2$ is mainly $p_x+p_y$-like. Therefore, in the case of a non-linear interpolation of the $\mathbf{k}\cdot\mathbf{p}$ parameters, the degree of optical polarization will be determined by the orbital character in the valence bands VB$_1$ and VB$_2$. Looking at the effective masses of the different valence bands near the $\Gamma$ point and along the direction of the $A$-point, we find that VB$_1$ ($m_{\text{VB}_1}=0.23$ m$_0$) has a low effective mass, while VB$_2$ and VB$_3$ ($m_{\text{VB}_2}=4.03$ m$_0$, $m_{\text{VB}_3}=1.14$ m$_0$) exhibit high effective masses. However, given the smaller crystal field splitting energy in the case of a nonlinear interpolation of the $\mathbf{k}\cdot\mathbf{p}$ parameters, quantum confinement in a $c$-plane Al$_{0.50}$Ga$_{0.50}$N QW may now give rise to strong band mixing effects as VB$_1$, VB$_2$ and VB$_3$ may be energetically very close and QW states are then formed from all three bands as linear combinations. As such, the degree of optical polarization may be strongly impacted by this effect and may differ from the results predicted when linearly interpolated parameters are used to calculate the QW electronic structure.

\section{Conclusion}

In summary, we have investigated the composition dependence of $\mathbf{k}\cdot\mathbf{p}$ parameters for wurtzite (Al,Ga)N alloys. These parameters are key ingredients for widely available theoretical and simulation models to analyze electronic, optical and carrier transport properties of (Al,Ga)N-based heterostructures and devices. 

To gain insight into the composition dependence of the $\mathbf{k}\cdot\mathbf{p}$ band parameters, we use GGA-DFT calculations to sample the electronic structure of (Al,Ga)N alloys over the full composition range. The DFT data were fitted with a $\mathbf{k}\cdot\mathbf{p}$ model widely used in the literature. Thus, the obtained $\mathbf{k}\cdot\mathbf{p}$ parameter set is compatible with, for instance, commercially available software packages, which, in general, do not allow to define "arbitrary" $\mathbf{k}\cdot\mathbf{p}$ Hamiltonians.

Our investigations show that the extracted band parameters for GaN and AlN agree well with literature data. However, our study also reveals that the widely used approximation of a linear interpolation of the GaN and AlN parameter sets with composition breaks down in (Al,Ga)N alloys. We find that to describe the DFT data accurately, composition-dependent bowing parameters are required for most of the (Al,Ga)N $\mathbf{k}\cdot\mathbf{p}$ parameters. We provide here an empirical expression to predict (Al,Ga)N $\mathbf{k}\cdot\mathbf{p}$ band parameters at desired alloy compositions. 

Moreover, we furnish initial insight into consequences of the nonlinear behavior of the $\mathbf{k}\cdot\mathbf{p}$ parameters with Al content for the bulk band structure of (Al,Ga)N alloys. We observed, for example, that band-ordering and energetic separation of the different valence bands are strongly impacted by the nonlinear evolution of the $\mathbf{k}\cdot\mathbf{p}$ parameter set with composition. The obtained results indicate that optical properties in (Al,Ga)N bulk alloys but also QW systems can differ significantly depending on whether a linear or nonlinear interpolation scheme to determine the $\mathbf{k}\cdot\mathbf{p}$ parameters is used.

\section*{Appendix A}
The \textbf{k}$\cdot$\textbf{p} Hamiltonian used in the present work is based on the work of Ghersoni \emph{et al.}~\cite{gershoni1993calculating}, and is also described in Refs.~\cite{chuang1996k, winkelnkemper2006interrelation}. For a wurtzite structure the $8\times8$ Hamiltonian in the basis  (\textbar\ensuremath{S}\textuparrow \textrangle, \textbar\ensuremath{X}\textuparrow \textrangle, \textbar\ensuremath{Y}\textuparrow \textrangle, \textbar\ensuremath{Z}\textuparrow \textrangle, \textbar\ensuremath{S}\textdownarrow\textrangle, \textbar\ensuremath{X}\textdownarrow\textrangle, \textbar\ensuremath{Y}\textdownarrow \textrangle, \textbar\ensuremath{Z}\textdownarrow\textrangle) is described in matrix form as:

\begin{equation}
\hat{H}=\left[\begin{array}{cc}
G(\mathbf{k}) & \Gamma\\
-\Gamma^{*} & G^{*}(\mathbf{k})
\end{array}\right]\,\, .
\label{eq:Hamiltonian}
\end{equation}
Here, $G(\mathbf{k})$ and  $\Gamma$ are  $4\times4$ matrices and $G(\mathbf{k})$ can be further decomposed into $4\times4 $ matrices:
\begin{equation}
 G(\mathbf{k})=G_{1}(\mathbf{k})+G_{2}(\mathbf{k})+G_\text{sp}+G_\text{cf}\,\, .
\end{equation}
The matrix $G_{1}(\mathbf{k})$ is given by
\begin{equation}
G_{1}(\mathbf{k})=\left[\begin{array}{cccc}
E_{c} & iP_2k_x & iP_2k_y & iP_1k_z\\
-iP_2k_x & E_{v} & 0 & 0\\
-iP_2k_y & 0 & E_{v} & 0\\
-iP_1k_z & 0 & 0 & E_{v}
\end{array}\right]\,\, ,
\end{equation}
where $E_c$ and $E_v$ are the conduction and valence band edge, respectively. $G_{2}(\mathbf{k})$ has the following form:
\begin{widetext}
\begin{equation}
G_{2}(\mathbf{k})=\left[\begin{array}{cccc}
A_2^{'}(k_x^{2}+k_y^2) + A_1^{'}k_z^{2} & B_2k_{y}k_{z} & B_2k_{x}k_{z} & B_1k_{x}k_{y}\\
B_2k_{y}k_{z} & L_{1}^{'}k_{x}^{2}+M_{1}k_{y}^{2}+M_{2}k_{z}^{2} & N_{1}^{'}k_{x}k_{y} & N_{2}^{'}k_{x}k_{z}-N_{3}^{'}k_{x}\\
B_2k_{x}k_{z} & N_{1}^{'}k_{x}k_{y} & M_{1}k_{x}^{2}+L_{1}^{'}k_{y}^{2}+M_{2}k_{z}^{2} & N_{2}^{'}k_{y}k_{z}-N_{3}^{'}k_{y}\\
B_1k_{x}k_{y} & N_{2}^{'}k_{x}k_{z}+N_{3}^{'}k_{x} & N_{2}^{'}k_{y}k_{z}+N_{3}^{'}k_{y} & M_{3}k_{x}^{2}+M_{3}k_{y}^{2} +L_{2}^{'}k_{z}^{2}\\
\end{array}\right]\,\, ,
\end{equation}
\end{widetext}
where terms $L_{i}^{'}$, $N_{i}^{'}$ and $M_{i}$ are associated with the Luttinger-like parameters $A_{i}$ with $i=1....6$, while the terms $A'_1$ and $A'_2$ are determined by conduction band effective masses $m_{e}^{\bot}$ and $m_{e}^{\Vert}$, respectively:  
\begin{subequations} \label{eq:eq11}

\begin{equation}
\begin{split}
A_1^{'}=\frac{\hbar^{2}}{2m_{e}^{\Vert}}\\
A_2^{'}=\frac{\hbar^{2}}{2m_{e}^{\bot}}\\
\end{split}
\end{equation}
\begin{equation}
\begin{split}
&L_{1}^{'}=\frac{\hbar^{2}}{2m_{0}}\left(A_{2}+A_{4}+A_{5}\right)\,\, , \\
&L_{2}^{'}=\frac{\hbar^{2}}{2m_{0}}\left(A_{1}\right)\,\, , \\
\end{split}
\end{equation}

\begin{equation}
\begin{split}
&M_{1}=\frac{\hbar^{2}}{2m_{0}}\left(A_{2}+A_{4}-A_{5}\right)\,\, , \\
&M_{2}=\frac{\hbar^{2}}{2m_{0}}\left(A_{1}+A_{3}\right)\,\, , \\
\end{split}
\end{equation}

\begin{equation}
\begin{split}
&N_{1}^{'}=\frac{\hbar^{2}}{2m_{0}}\left(2A_{5}\right)\,\, , \\
&N_{2}^{'}=\frac{\hbar^{2}}{2m_{0}}\left(\sqrt{2}A_{6}\right)\,\, , \\
&N_{3}^{'}=i\left(\sqrt{2}A_{7}\right)\,\, . \\
\end{split}
\end{equation}
\end{subequations}
Following Ghersoni \emph{et al.}~\cite{gershoni1993calculating}, we set the $B$  parameter to zero. This is consistent with the work of Rinke \emph{et al.}~\cite{rinke2008consistent} where it is discussed that including this term did not improve the fitting of the band structures. Moreover, the $A_{7}$ parameter is set to zero in our calculations, which is a widely used approximation and makes the model compatible with, e.g., $\mathbf{k}\cdot\mathbf{p}$ Hamiltonians underlying commercially available software packages such as nextnano~\cite{birner2007nextnano} and Crosslight APSYS~\cite{crosslight2024}. Furthermore, the parameters $P_{1/2}$ that describe conduction band valence band coupling are set to zero, which is also a widely used approximation for wide-band gap materials such as (Al,Ga)N alloys~\cite{fonoberov2003excitonic, fu2011exploring, gladysiewicz2019material}. The reduced form of the Hamiltonian is often referred to as a $6+2$ band model.

In the standard definition of the Hamiltonian, Eq.~(\ref{eq:Hamiltonian}), the matrices $G_\text{sp}$ and $\Gamma$ describe the spin-orbit coupling. As discussed in the main text, we use these terms here to account for alloy disorder-induced band mixing and lifting of degeneracies in the band structure of (Al,Ga)N alloys. In general, we use the standard definitions of $G_\text{sp}$ and $\Gamma$ in our fitting procedure: 
\begin{equation}
G_\text{sp}=\frac{\Delta_\text{sp}}{3}\left(\begin{array}{cccc}
0 & 0 & 0 & 0\\
0 & 0 & -i & 0\\
0 & i & 0 & 0\\
0 & 0 & 0 & 0
\end{array}\right)\,\, ,
\end{equation}
and
\begin{equation}
\varGamma=\frac{\Delta_\text{sp}}{3}\left(\begin{array}{cccc}
0 & 0 & 0 &  0 \\
0 & 0 & 0 &  1\\
0 & 0 & 0 & -i\\
0 & -1 & i & 0
\end{array}\right)\,\, .
\end{equation}
Finally, the crystal field splitting contribution, present in the wurtzite crystal structure, is described by:
\begin{equation}
G_\text{cf}=\Delta_\text{cf}\left(\begin{array}{cccc}
0 & 0 & 0 & 0\\
0 & 1 & 0 & 0\\
0 & 0 & 1 & 0\\
0 & 0 & 0 & 0
\end{array}\right)\,\, .
\end{equation}

\section*{Appendix B}
Expressions for the effective hole effective masses parallel and perpendicular to the wurtzite $c$-axis have been given in Refs.~\cite{rinke2008consistent, chuang1996k} in terms of the Luttinger-like parameters $A_i$ with $i=1...6$. Following Refs.~\cite{rinke2008consistent, chuang1996k} the masses can be calculated close to or far from the $\Gamma$-point, and the corresponding equations are summarized below.

\paragraph{Hole masses close to $\Gamma$-point}
\begin{subequations} \label{eq:eq17}
\begin{equation}
\begin{split}
&m_{0}/m_{\text{VB}_1}^{\Vert}	=	-(A_{1}+A_{3}) \\
&m_{0}/m_{\text{VB}_1}^{\bot}	=	-(A_{2}+A_{4}) \\
\end{split}
\end{equation}

\begin{equation}
\begin{split}
&m_{0}/m_{\text{VB}_2}^{\Vert}	=	-\left[A_{1}+\left(\frac{E_{\text{VB}_2}}{E_{\text{VB}_2}-E_{\text{VB}_3}}\right)A_{3}\right] \\
&m_{0}/m_{\text{VB}_2}^{\bot}	=	-\left[A_{2}+\left(\frac{E_{\text{VB}_2}}{E_{\text{VB}_2}-E_{\text{VB}_3}}\right)A_{4}\right] \\
\end{split}
\end{equation}

\begin{equation}
\begin{split}
&m_{0}/m_{\text{VB}_3}^{\Vert}	=	-\left[A_{1}+\left(\frac{E_{\text{VB}_3}}{E_{\text{VB}_3}-E_{\text{VB}_2}}\right)A_{3}\right] \\
&m_{0}/m_{\text{VB}_3}^{\bot}	=	-\left[A_{2}+\left(\frac{E_{\text{VB}_3}}{E_{\text{VB}_3}-E_{\text{VB}_2}}\right)A_{4}\right] \\
\end{split}
\end{equation}
\end{subequations}
Here, $E_{\text{VB}_2}$ and $E_{\text{VB}_3}$ are the energies of the three valence bands at the $\Gamma$-point described by Eqs.~(\ref{eq:VBE2}) and~(\ref{eq:VBE3}) in Sec.~\ref{sec:AlGaN_kp_parameters}.\\

\paragraph{Hole masses far from $\Gamma$-point}

\begin{subequations} \label{eq:eq18}
\begin{equation}
\begin{split}
&m_{0}/m_{\text{VB}_1}^{\Vert}	=	-(A_{1}+A_{3}) \\
&m_{0}/m_{\text{VB}_1}^{\bot}	=	-(A_{2}+A_{4} -A_{5}) \\
\end{split}
\end{equation}

\begin{equation}
\begin{split}
&m_{0}/m_{\text{VB}_2}^{\Vert}	=	-(A_{1}+A_{3}) \\
&m_{0}/m_{\text{VB}_2}^{\bot}	=	-(A_{2}+A_{4} + A_{5}) \\
\end{split}
\end{equation}

\begin{equation}
\begin{split}
&m_{0}/m_{\text{VB}_3}^{\Vert}	=	-A_{1} \\
&m_{0}/m_{\text{VB}_3}^{\bot}	=	-A_{2} \\
\end{split}
\end{equation}
\end{subequations}

\section*{Acknowledgments}
This work was financially supported by the Marie Sklodowska-Curie Actions programme (EU) and Taighde {\'E}ireann -- Research Ireland, formerly Science Foundation Ireland (SFI), co-fund programme SPARKLE (Grant No. 847652), by Research Ireland (Grant No. 21/FFP-A/9014), and by the Research Ireland-funded Irish Photonic Integration Centre (IPIC; Grant No. 12/RC/2276\_P2). ams-OSRAM International GmbH gratefully acknowledges support by the Federal Ministry for Economic Affairs and Climate Action on the basis of a decision by the German Bundestag and by the Bavarian Ministry of State for Economic Affairs, Regional Development and Energy as well as by the European Union – NextGenerationEU for the Important Project of Common European Interest on microelectronics and communication technologies (IPCEI ME/CT) - OptoSuRe (16IPCEI221).

\section*{Data Availabilty}
The data that support the findings of this study are available from the corresponding author upon reasonable
request.

\bibliographystyle{unsrtnat}
\bibliography{database}

\end{document}